%% file: sample-acmsmall-submission.tex
\documentclass[acmsmall]{acmart}

\setcopyright{none}
\renewcommand\footnotetextcopyrightpermission[1]{}

\AtBeginDocument{%
  }

\usepackage[normalem]{ulem}
\newcommand{\smallsection}[1]{\smallskip\noindent\textbf{#1}}
\usepackage{soul}
\usepackage{dirtytalk}
\usepackage{subcaption}
\usepackage{caption}
\usepackage[dvipsnames]{xcolor}

\begin{document}

\title{Machine Learning Practitioners’ Views on Data Quality in Light of EU Regulatory Requirements: A European Online Survey}

\author{Yichun Wang}

\orcid{1234-5678-9012}
\affiliation{%
  \institution{University of Amsterdam}
  \country{The Netherlands}}
\authornote{Yichun Wang designed the study, conducted the survey, and drafted the paper under the supervision of the other authors. Kristina Irion contributed to the study design, legal analysis, and reviewing and provided supervision on the legal aspects of the work. Paul Groth contributed to the study design and reviewing and provided supervision throughout the project. Hazar Harmouch managed the collaboration and contributed to the study design, reviewing and refining the draft.}

\author{Kristina Irion}
\affiliation{%
 \institution{University of Amsterdam}
 \country{The Netherlands}}
 

\author{Paul Groth}
\affiliation{%
  \institution{University of Amsterdam}
  \country{The Netherlands}
}

\author{Hazar Harmouch}
\affiliation{%
 \institution{University of Amsterdam}
 \country{The Netherlands}}

\renewcommand{\shortauthors}{Wang et al.}

\input{01-Abstract}




\maketitle


\input{02-Introduction}
\input{03-research_methodology}
\input{04-literature_study}

\input{05-Survey}
\input{06-Results_Discussion}

\input{07-Recommendations}
\input{08-Conclusion_Future}

\section*{Artifacts}
Survey questions and additional survey results are provided in the supplementary materials.

\section*{Acknowledgments}
The authors thank all survey participants for their time, candour, and willingness to share their experiences and insights. This research was performed in the context of the AI4FinTech initiative of the University of Amsterdam.

\bibliographystyle{ACM-Reference-Format}
\bibliography{sample-base}

\appendix

\end{document}

%% file: 01-Abstract.tex
\begin{abstract}
Understanding how data quality aligns with regulatory requirements in machine learning (ML) systems presents a critical challenge for practitioners navigating the evolving EU regulatory landscape. To address this, we first propose a practical framework aligning established data quality dimensions with specific EU regulatory requirements. Second, we conducted a comprehensive online survey with over 180 EU-based data practitioners, investigating their approaches, key challenges, and unmet needs when ensuring data quality in ML systems that align with regulatory requirements. Our findings highlight crucial gaps between current practices and regulatory expectations, underscoring practitioners' need for more integrated data quality tools and better collaboration between technical and legal practitioners. These insights inform recommendations for bridging technical expertise and regulatory compliance, ultimately fostering responsible and trustworthy ML deployments.
\end{abstract}

%% file: 02-Introduction.tex
\section{Data quality and compliance from practitioners' perspectives}


The importance of data in the development of ML systems cannot be overstated \cite{datamanagementchallenges,schelter2015challenges,stoyanovich2022responsible,karlavs2024navigating}. In real-world scenarios, the lifecycle of data within ML systems typically encompasses several key stages: initial data collection, preprocessing, model training, and ongoing post-deployment monitoring \cite{priestley2023survey, rangineni2023analysis, 10.1145/3453444}. This lifecycle is rarely a linear process, but rather a continuous and iterative cycle, allowing for the regular updating and refinement of models and datasets. Multiple professionals and teams collaboratively execute this lifecycle. Data collectors and labellers ensure the accuracy and relevance of data. Product design and management professionals contribute essential strategic direction. Engineers dedicate significant time to the often underappreciated tasks of data cleaning and preparation. Data scientists and ML engineers transform processed datasets into functioning ML models. Together, this diverse ecosystem of practitioners transforms input data into valuable information and products.

Although data quality is critically important, proactive steps to safeguard it often receive far less attention than algorithm development and model optimisation~\cite{sambasivan2021everyone}. The data management community has extensively explored various data quality issues~\cite{liu2016rethinking, xu2002data, redman2001data,grafberger2022data,guha2024automated,karlavs2023data}. However, practitioners continue to face substantial challenges in ensuring high data quality throughout the ML lifecycle \cite{cai2015challenges, whang2020data, whang2023data}. 

\smallsection{Regulatory compliance.} Practitioners working with data-driven systems encounter not only technical data quality challenges, but also the need to ensure compliance with relevant legal requirements \cite{labadie2023building}. This is further accelerated by the prevalence of European legal frameworks, which add new complexities related to data quality in ML \cite{hackerLegalFrameworkAI2021}. The European Union (EU) has established a legislative framework governing the use of data for ML applications. The General Data Protection Regulation (GDPR) \cite{GDPR2016} applies when personal data is part of the dataset. The recently adopted Artificial Intelligence Act (AI Act) \cite{AIA2024} formulates data governance requirements for high-risk AI systems.

Data practitioners need to apply legal texts and translate them into effective, actionable data management practices~\cite{hackerLegalFrameworkAI2021}. Regulations such as the GDPR and the AI Act leave practitioners with open questions about how to best operationalise them. For instance, Art.~5(1)(d) of the GDPR mandates that personal data be \say{accurate and, where necessary, kept up to date}, yet translating this requirement into actionable data quality measurements is far from straightforward. By acknowledging this interpretative flexibility, we see an opportunity to help practitioners develop concrete understandings that align with both technical needs and regulatory mandates.


\smallsection{An existing gap.} 
The academic and industry discourse has explored two aspects separately: technical strategies such as automated data cleaning \cite{10.1145/2882903.2912574, gudivada2017data}, data validation frameworks \cite{MOODY2003619,MLSYS2019_928f1160,schelter2018automating}, and ongoing monitoring systems \cite{Du2020MissingDP, Yaqoob2021BlockchainFH}; and compliance-focused strategies such as regulatory assessment tools and legal checklists \cite{IBM_AI_Data_Management, hackerLegalFrameworkAI2021, guldimannCOMPLAIFrameworkTechnical2024a}.These two areas are often discussed in isolation from each other, lacking a practical, integrated approach that effectively combines technical expertise with regulatory compliance throughout the ML lifecycle.
In addition, there is a potential disconnect between academic research and the practical realities faced by industry practitioners. 
Although much of current research prioritises technical solutions, the actual implementation of these solutions in real-world scenarios often faces significant challenges, including the absence of standardised frameworks and widely accepted definitions (e.g., varying interpretations of ``fairness'' and ``transparency''), and the varying levels of regulatory awareness and familiarity among practitioners or in different organisational contexts \cite{Holstein2018ImprovingFI}. In domains with higher stakes, such as healthcare, finance, or public services, these problems are more prevalent~\cite{Syed2023DigitalHD, AlOkaily2024FinancialDM}. 
Consequently, there remains an underexplored area concerning how data quality management can be effectively aligned with regulatory requirements within real-world ML practices. A related interview study of EU practitioners provides qualitative evidence of these gaps in day-to-day work, and reports persistent difficulties in translating legal principles into engineering practice~\cite{wang2026detective}.


\smallsection{Practitioners at the centre.} We have chosen to focus on practitioners’ perspectives, which have been less explored concerning data-related regulatory compliance. This focus stems from the direct yet nuanced role that practitioners play in the execution of data management processes that ought to meet legal standards. While they may not bear legal liability for non-compliance, they operate on the front lines, directly handling data and implementing procedures to ensure compliance through their hands-on work. They are expected to be well-informed and possess the knowledge and ethical foundation to question and potentially refuse practices that could lead to non-compliant activities. Furthermore, effective collaboration between these practitioners and the legal and compliance teams within their organisations is highly desirable. Therefore, understanding the challenges they face is essential for identifying the practical barriers to compliance. 

To gain a comprehensive, enterprise-wide perspective on how compliance requirements intersect with technical standards, we investigate the direct experiences of practitioners. The novelty of this research lies in its integration of theoretical, regulatory, and empirical insights to offer guidance on ensuring data quality that aligns with EU regulations. By leveraging the valuable input of over 180 EU data practitioners through surveys, our study provides a robust empirical foundation to inform academia, industry, and policymakers. The findings will contribute to the development of actionable insights and strategies that promote more effective, regulation-aligned data management. This alignment, in turn, will enable organisations and practitioners to better balance compliance demands with operational efficiency and innovation, ultimately fostering trust and facilitating the wider adoption of AI technologies.

\smallsection{Contribution.}
In summary, our research makes the following main contributions:
\begin{enumerate}
\item A comprehensive framework that intersects established data quality dimensions from technical literature and specific regulatory requirements in the EU. By doing so, we provide practitioners with a practical vocabulary to translate regulatory mandates into clear, actionable data quality practices.
\item Empirical insights from a survey of over 180 EU data practitioners investigating current approaches, challenges, and needs in ensuring data quality compliant with EU regulatory standards.
\item Lessons from our empirical investigation highlight the multifaceted nature of data quality dimensions, particularly in regulated ML contexts. This includes strategic approaches to balance technical excellence with regulatory demands throughout the ML lifecycle.
\end{enumerate}

\smallsection{Outline.}
The remainder of the paper is structured as follows. Section~\ref{sec:researchquestion}  introduces our research objectives, research question, and methodology. Section~\ref{sec:related_work} reviews relevant literature from both technical and regulatory perspectives, leading to the proposed integrated vocabulary of data quality dimensions aligned with EU regulations. Sections~\ref{sec:surveymethodology} and \ref{sec:results} detail our survey methodology and the empirical findings. Finally, Sections~\ref{sec:recommendations} and \ref{sec:conclusion} conclude with a summary of contributions, limitations, and directions for future research.

%% file: 03-research_methodology.tex
\section{Research objectives and methodology}
\label{sec:researchquestion} 
In this study, we aim to empirically investigate how data practitioners approach the critical task of managing data quality while adhering to legal requirements stemming from EU law. We seek to uncover the practical challenges and unmet needs faced by these data practitioners in aligning their data management strategies with regulatory mandates. By integrating insights derived from a comprehensive review of technical literature, an analysis of relevant regulatory texts,  and the findings of a survey involving over 180 data practitioners across the EU, this research identifies key discrepancies and limitations between theory and current practice. Ultimately, the goal is to provide actionable strategies that can facilitate a more effective alignment of data quality management practices with EU regulatory requirements. More specifically, this study seeks to address the following primary research question:  
\begin{center}
\emph{How do technical data quality dimensions align with regulatory requirements, and how do practitioners currently address challenges and unmet needs when ensuring data quality in ML systems subject to these regulations?}
\end{center}

To effectively address this central question, the study will delve into the following specific sub-questions: 
\begin{itemize}
\item {SRQ1: How to interpret and operationalise definitions of data quality to ensure alignment with regulatory requirements?}

\item {SRQ2: How do practitioners prioritise data quality aspects and adopt best practices throughout the ML lifecycle to align with EU regulatory standards?}

\item {SRQ3: What tools and collaboration dynamics do practitioners currently use to manage data quality in alignment with regulatory requirements, and what gaps remain in their approaches?}
\end{itemize}

To address these research questions, we adopt a design science research approach that integrates both theoretical and empirical aspects of ensuring data quality in alignment with EU regulatory requirements~\cite{Shull2007GuideTA}. This approach encompasses \textit{two} key components: (1) {\itshape Framework Development:} we establish a robust framework that synthesises data quality literature and EU law requirements relevant to (personal) data. By identifying the most significant overlaps in data quality dimensions from both perspectives, we define a clear operational scope for our following investigation; (2) {\itshape Survey:} based on the framework, we conducted a large-scale survey involving over 180 data practitioners across the EU to gather broad empirical data on the challenges and practices related to data quality and regulatory compliance. Participants in the survey were carefully selected to represent a diverse range of domains and roles, ensuring a comprehensive perspective on how data quality practices are shaped by regulatory contexts. 



The target audience of this study includes data practitioners and researchers who want to understand legal compliance requirements in ML systems regarding data processing, as well as informatics legal professionals such as compliance officers seeking deeper technical perspectives to ensure compliance. By discussing existing issues in current data quality practices and desired solutions, we aim to motivate further research in this area. We believe that addressing these data-related challenges requires collaboration between diverse communities, which includes both technical and legal sides. 

%% file: 04-literature_study.tex
\section{Literature review}
\label{sec:related_work}
In this section, we review relevant literature to establish a foundation for understanding data quality in the context of ML systems and regulatory requirements. We first examine how data quality has been conceptualised in technical domains, focusing on multidimensional frameworks and their application throughout the ML lifecycle. We then explore selected EU regulations (particularly the GDPR and AI Act) to identify key provisions that impact data quality management practices. Finally, we analyse the intersection between these technical and legal perspectives, highlighting both alignments and tensions that practitioners must navigate when implementing data quality measures that satisfy regulatory demands.

\subsection{Data quality \& ML: the technical perspective}

The widespread adoption of ML across numerous domains is intrinsically linked to the availability of vast amounts of data. However, a common misconception is that more data inherently leads to better outcomes \cite{jain2020overview}. For example, there has been a tendency to collect extensive datasets without adequate attention to their quality in response to the rapid development of large language models (LLMs). This oversight is problematic, as the efficiency and reliability of ML models are fundamentally dependent on data quality \cite{rangineni2023analysis}. 

The adage ``garbage in, garbage out'' (GIGO) aptly encapsulates the critical influence of data quality on ML systems \cite{kilkenny2018data}. Poor data quality can lead to inaccurate predictions \cite{budach2022effects}, biased outcomes \cite{EWAF23}, and ultimately diminish user trust. Furthermore, the significant time and effort that data practitioners typically spend on data cleaning and preparation underscore the practical challenges of ensuring high data quality \cite{munson2012study}. Studies have indicated that 40\% to 60\% of practitioners’ time is devoted to these tasks, highlighting the urgency for more efficient, automated, and scalable data quality management solutions \cite{anacondareport}.

\smallsection{Data quality frameworks.} Research on data quality has evolved significantly since the 1980s~\cite{chrisman1983role, wang1996beyond}. Data quality is now widely understood as a multidimensional concept that reflects the socio-technical complexity of contemporary data ecosystems. Numerous frameworks and classifications of data quality dimensions have been proposed over time~\cite{wang1998product, lee2002aimq, pipino2002data}. A foundational and widely adopted framework by Wang and Strong categorises data quality into four broad categories: intrinsic, contextual, representational, and accessibility~\cite{wang1996beyond}. Within these categories, various specific dimensions emerge. For example, intrinsic data quality refers to the extent data accurately represents true or actual values, emphasising accuracy and traceability. Contextual quality addresses whether data is relevant and sufficient for the user's specific tasks, encompassing dimensions like relevance, timeliness, and completeness, among others. Representational data quality relates to how clearly and intelligibly data is presented, often supported by detailed documentation practices in ML. Accessibility focuses on ensuring data is readily available, retrievable, and secure. In practice, stakeholders working with sensitive data balance improved accessibility with security needs.

Beyond technical and operational factors, recent work emphasises that poor data quality cannot be decoupled from its societal and ethical implications, including fairness, privacy, and societal trust \cite{pitoura2020social, Pessach2022ARO, Hort2022BiasMF, jain2016big}. For instance, bias in ML systems is often a direct consequence of poor data quality, particularly when the training data is biased or unrepresentative of the target population \cite{crowdsourcingfairness}. If the data used to train a model reflects existing societal prejudices or historical inequalities, the model is likely to reinforce and even amplify these biases in its predictions, resulting in unfair or discriminatory treatment of certain groups. Therefore, fairness (avoiding biases that disadvantage specific groups) is now recognised as integral to data quality in contexts like algorithmic decision-making \cite{floridi2018ai4people, zook2017ten}. Furthermore, inadequate data management practices can result in privacy breaches, exposing sensitive information and violating data protection standards such as the GDPR, thereby diminishing user confidence and stakeholder trust \cite{mittelstadt2019explaining, janssen2020data}. Thus, privacy (ensuring compliance with data protection regulations like the GDPR) and accountability (traceability of data misuse) are increasingly framed as quality criteria, particularly in sensitive domains such as healthcare and public policy \cite{schwabe2024metric, mittelstadt2017individual}. 

\smallsection{Prioritisation.} While these dimensions are all crucial, their prioritisation and specific interpretation by practitioners can differ in the context of ML development \cite{priestley2023survey}.  A common understanding is the notion of ``fitness for use'', emphasising that data quality is inherently context-dependent and should be evaluated based on the specific requirements of the task at hand \cite{examiningdataquality}. For example, an ML system designed for medical diagnosis might prioritise accuracy and completeness to ensure patient safety and effective treatment \cite{ng2023perceptions}, while a recommendation system might place greater emphasis on timeliness and relevance to provide users with up-to-date and personalised suggestions \cite{ko2022survey}. Consequently, understanding data quality requires careful consideration of context-specific needs, particularly in ML settings, where irrelevant or poorly selected features can degrade model performance \cite{budach2022effects}. 

Such prioritisation themes reveal a persistent gap between theoretical frameworks and practice. Technical and operational criteria dominate due to their direct link to short-term model performance or operational efficiency, while ethical dimensions are often treated as optional ``add-ons'' unless mandated by regulation. However, this imbalance carries significant risks, for instance, biased training data perpetuates inequitable ML outcomes \cite{buolamwini2018gender}, while poor documentation undermines transparency and reproducibility \cite{pineau2021improving, gundersen2018state}.

\smallsection{ML Lifecycle.} 
The iterative nature of the ML lifecycle means data quality must be actively addressed at every stage, from initial data collection and preprocessing to model training, deployment, and ongoing monitoring \cite{gudivada2017data}. Data collection and labelling often emerge as significant bottlenecks in deploying ML applications \cite{huang2024data}. For instance, a facial recognition system trained on non-representative data will struggle to generalise across demographics \cite{howard2017addressing}. Operational dimensions like documentation (e.g., metadata standards) and traceability (e.g., data lineage tracking) are often underprioritised at this stage. The data preparation stage addresses issues such as missing values, duplicates, inconsistencies, and outliers \cite{krishnan2016activeclean}. Techniques like imputation, removal of duplicates, and normalisation transform the data into a suitable format for ML algorithms \cite{ahsan2021effect}. The operational rigour of documentation is crucial here to interpret data semantics accurately. For example, a missing ``0'' in a medical dataset could represent a healthy measurement or a data entry oversight \cite{cismondi2013missing}. 

Selecting relevant features based on quality assessments contributes to building parsimonious and effective models. However, even pristine data can be corrupted by poorly designed transformations or biased feature choices. For instance, aggregating transactional data into weekly averages might obscure critical temporal patterns (contextual relevancy), leading models to miss fraud signals tied to real-time spikes. Moreover, feature selection often reflects implicit assumptions: a feature deemed “relevant” during engineering may inadvertently exclude marginalised groups or overemphasise spurious correlations \cite{fenza2021data}. A healthcare model predicting patient readmissions, for example, might prioritise features like prior hospitalisation frequency while ignoring social determinants of health (e.g., access to transportation), perpetuating inequities in care \cite{cross2024bias}. 

The consequences of these interdependencies crystallise during model training and evaluation, where both technical and ethical dimensions face existential tests. For example, a healthcare ML system trained on accurate, complete data may still fail if the data lacks relevance to marginalised populations, such as excluding rural patients in a diabetes prediction model. Validation metrics like F1-scores or AUC-ROC provide narrow, technical assurances but often ignore operational and ethical dimensions, such as whether the model’s false negatives disproportionately harm vulnerable groups \cite{51764}. This disconnect reveals a systemic blind spot: teams optimise for measurable technical criteria while deprioritising harder-to-quantify ethical dimensions, storing up risks for later stages. 

These risks become pronounced during deployment and monitoring, where all dimensions collide in real-world contexts.  Operational dimensions like timeliness (e.g., detecting data drift) and traceability (e.g., auditing model decisions) become critical as models interact with dynamic, open-world data. A fraud detection system trained on historical transaction patterns may degrade in effectiveness as criminals adapt over time \cite{sudjianto2010statistical}. This decline in contextual relevance necessitates continuous updates to the training data. Meanwhile, ethical dimensions like privacy face new threats: edge devices collecting user data for retraining must balance completeness (technical) with GDPR compliance (legal). Yet, interventions here are inherently reactive. Practitioners may deploy drift detection algorithms or fairness-aware retraining, but these fixes are costlier and less effective than proactive dimension alignment in earlier stages. For instance, retroactively de-biasing a model is far more complex than curating representative training data upfront. 

In conclusion, data quality is integral to every stage of the ML lifecycle. While quality issues often become most apparent in the later stages, they are also the most challenging to address at that point. The key lies in iterative, stage-specific, and dimension-aware governance.

\subsection{Selected EU regulation of data used in ML}

We now review two key EU regulatory frameworks that shape how data is managed in ML systems, namely the GDPR and the AI Act. This analysis specifically focuses on public law regulations within the EU, excluding private law such as liability under contract or tort law~\cite{hackerLegalFrameworkAI2021}. The scope is deliberately limited to regulations governing data used throughout the ML lifecycle, rather than addressing regulations that primarily concern ML outcomes, such as anti-discrimination law. Although intellectual property laws regarding the legality and lawfulness of using third-party data for ML apply, these matters are normally handled by legal departments rather than data practitioners.

\smallsection{General Data Protection Regulation (GDPR).}
The GDPR represents a significant milestone in data protection law in the EU~\cite{Hoofnagle02012019}. The regulation was formally adopted in April 2016 and became enforceable on May 25, 2018. This regulation governs the processing of personal data, which includes any information relating to an identified or identifiable natural person (Art.~4(1))~\cite{DistinguishingPersonal}. Processing itself includes virtually any operation performed on personal data, from collection and storage to analysis and deletion (Art.~4(2)). Crucially, the GDPR has a broad territorial scope~\cite{Voigt2017}. It applies not only to organisations established within the EU, but also to entities outside the Union that process EU residents' data in the context of offering them goods or services, or monitoring their behaviour (Art.~3).  The primary entities addressed by the regulation are ‘controllers’, who,  according to Art.~4(7) of the GDPR determine the purposes and means of processing, and ‘processors’, who, according to Art.~4(8) of the GDPR process data on the controller's behalf. In this paper, we explore the GDPR principles and requirements specifically in the context of ML systems.

At its core, the GDPR establishes fundamental principles in Art.~5 that govern all personal data processing activities~\cite{goddard2017eu, Hoofnagle02012019}. These principles mandate that data processing must be lawful, fair, and transparent; collected for specified, explicit purposes (purpose limitation); adequate, relevant, and limited to what is necessary (data minimisation); and accurate and kept up to date (accuracy and timeliness). Further reinforcing the requirement for lawful handling, Art.~6 stipulates that all personal data processing must have a valid legal basis, such as performance of a contract, explicit consent or legitimate interest~\cite{Hoofnagle02012019}. Moreover, the GDPR empowers individuals with specific rights over their data, including rights of access (Art.~15), rectification of inaccuracies (Art.~16), and erasure (`right to be forgotten', Art.~17). Correspondingly, these rights impose substantial obligations on data controllers to ensure compliance, facilitate the exercise of these rights, and ultimately maintain the quality and integrity of the personal data they manage~\cite{sirur2018we}.

\smallsection{The Artificial Intelligence Act (AI Act).}
Without prejudice to the GDPR, the EU introduced the AI Act in 2024. This regulation aims to establish harmonised rules for the development, placing on the market, and use of AI systems (Art.~1). Its primary goals are to promote trustworthy AI while balancing innovation with fundamental rights protection~\cite{laux2024trustworthy, smuha2021eu}. The AI Act entered into force in 2024, with its provisions becoming applicable progressively over the subsequent 24 to 36 months of the implementation period. In particular, the data governance obligations for high-risk AI systems under Art. 10 become applicable in August 2026, though this may be postponed pending adoption of the Digital Omnibus on AI. These obligations are not yet in force, but they already shape emerging discussions on data quality practices in ML systems. It distinguishes between developers who create AI systems and deployers who implement them in real-world contexts. Its scope covers providers placing AI systems on the EU market or affecting EU citizens, regardless of the provider's location (Art.~2(1)). It excludes any research, testing, or development of AI systems or models prior to their placement on the market or into service (Art.~2(8)). The Act defines an `AI system' as software developed using techniques like ML that can generate outputs such as content, predictions, recommendations, or decisions influencing environments (Art.~3(1)). Consequently, most of the ML systems discussed in this paper will likely qualify as `AI systems' under this legal framework.

The AI Act adopts a risk-based approach, with `high-risk' AI systems subject to stringent requirements~\cite{novelli2024taking}. High-risk AI systems include those used in critical infrastructures, education, employment, essential private and public services, law enforcement, migration management, and administration of justice. For these high-risk applications, Art.~10 mandates comprehensive data governance practices for high-risk AI systems, requiring training data to be relevant, representative, and free from errors~\cite{hackerLegalFrameworkAI2021}. Art.~13 imposes transparency obligations, compelling developers to provide clear documentation about system functionality to enable deployers to interpret the system's output and use it appropriately. Meanwhile, Art.~14 requires human oversight mechanisms to minimise risks. While these obligations, particularly Art.~10's focus on data quality, are highly relevant to this paper's themes, it is important to recognise that many ML systems discussed might not qualify as `high-risk' under the Act.

\subsection{Regulatory compliant data quality}
\label{sec:framework}

\begin{table*}
\caption{The Intersection of Data Quality Dimensions from Technical and Legal Perspectives}
\label{tab:data-quality}
\scriptsize
\begin{tabular}
{p{0.11\linewidth}p{0.19\linewidth}p{0.32\linewidth}p{0.28\linewidth}}
\toprule
\textbf{Data Quality Dimensions} & \textbf{Technical Interpretation} & \textbf{Provisions in the GDPR as applicable} & \textbf{Provisions in the AI Act as applicable}\\
\midrule
\textbf{Accuracy} & Data is correct, reliable, and free from errors. & Requires data accuracy; erase or rectify incorrect personal data -- \textbf{Art.\,5(1)(d)}, \textbf{Recital 39}\newline Right to rectification of inaccurate personal data -- \textbf{Art.\,16} & Datasets shall be error-free -- \textbf{Art.\,10(3)}\newline Datasets should be, to the best extent possible, free of errors -- \textbf{Recital 67} \\
\midrule
\textbf{Traceability} & Changes or updates of the data should be tracked. & Security measures including logs -- \textbf{Art.\,32} & Mandatory automated logging and record-keeping for traceability -- \textbf{Art.\,12(3)(b)\&(c)}, \textbf{Art.\,19} \\

\midrule
\textbf{Relevancy} & Data is appropriate and applicable to the intended task. & Personal data shall be adequate, relevant, and limited to what is necessary in relation to the purposes -- \textbf{Art.\,5(1)(c)}, \textbf{Recital 39} & Requires assessment of the dataset's suitability -- \textbf{Art.\,10(2)(e)}\newline Datasets must be relevant to the intended purpose -- \textbf{Art.\,10(3)} \\

\midrule
\textbf{Timeliness} & Data is up-to-date and available when needed. & Personal data must be kept up to date -- \textbf{Art.\,5(1)(d)} & \- \\
\midrule
\textbf{Completeness} & All necessary components of data are present and adequately recorded. & Adequate and relevant personal data for intended purposes -- \textbf{Art.\,5(1)(c)} & Datasets must be complete for the intended purpose -- \textbf{Art.\,10(3)}\newline Completeness to avoid biased or inaccurate AI outputs -- \textbf{Recital 67} \\
\midrule
\textbf{Documentation} & Clear and comprehensive information should describe the data and its management. & Accountability principle mandates documented processing activities -- \textbf{Art.\,5(2)}, \textbf{Art.\,35}, \textbf{Recital 82}\newline Requires maintaining records of processing activities -- \textbf{Art.\,30(1), (3)\&(4)} & High-risk AI systems shall be accompanied by instructions for the input data -- \textbf{Art.\,13(3)(b)(vi)}\newline Record-keeping of input data that led to a match -- \textbf{Art.\,12}\newline The technical documentation shall contain information on the data -- \textbf{Art.\,11}, \textbf{Art.\,18}, \textbf{Annex IV}, \textbf{Annex XI} \\
\midrule
\textbf{Accessibility} & Data can be accessed and used by authorised users. & Right of access for data subjects -- \textbf{Art.\,15}\newline Data portability in structured formats (facilitating data accessibility) -- \textbf{Art.\,20} & \- \\
\midrule
\textbf{Security} & Personal data should be protected from unauthorised disclosure or misuse. & Mandates integrity and confidentiality through robust safeguards -- \textbf{Art.\,5(1)(f)}\newline Privacy by design/default embedded in systems development -- \textbf{Art.\,25}\newline Requires appropriate technical and organisational measures to ensure a level of security appropriate to the risk -- \textbf{Art.\,32} & Privacy protection throughout AI lifecycle, adhering to the GDPR principles -- \textbf{Recital 69}\newline Incorporates privacy and data protection into its risk management and ethical frameworks for AI systems -- \textbf{Recital 27}\newline AI Act does not affect personal data privacy in the GDPR -- \textbf{Art.\,2(7)} \\
\midrule
\textbf{Bias mitigation} & Data is impartial and free from biases, and treats groups equitably. & Automated individual decision-making should not be based on special categories of personal data -- \textbf{Art.\,22(4)}, \textbf{Recital 71} & Data governance and management practices to detect, prevent and mitigate possible biases -- \textbf{Art.\,10(2)(f)\&(g)}\newline Datasets shall have appropriate statistical properties regarding persons or groups of persons -- \textbf{Art.\,10(3)}\newline Exceptionally process special categories of personal data for bias detection and correction -- \textbf{Art.\,10(5)}, \textbf{Recital 70}\newline Attention to bias mitigation in the data sets -- \textbf{Recital 67} \\

\bottomrule
\end{tabular}

\end{table*}

Here, we define ``Regulatory compliant data quality” as the degree to which data meets technical quality standards while adhering to applicable EU regulatory frameworks, such as the GDPR and the AI Act. This means data must not only be technically ``fit for use”, but also lawfully collected, processed, stored, and validated against legal requirements. Given the complexity and multidimensional nature of data quality, operationalising clear definitions that align with legal standards remains challenging. 
To address the first sub-research question (SRQ1), we systematically review key data quality dimensions widely recognised in data engineering and quality assurance literature~\cite{wang1996beyond, pipino2002data}. In Table~\ref{tab:data-quality}, we explicitly map these technical perspectives to the corresponding legal mandates in the GDPR and the AI Act. Through examining each dimension, we interpret how existing technical definitions can be operationalised within regulatory frameworks, as well as highlighting both synergies and potential tensions. This facilitates actionable guidance for practitioners to achieve regulatory-compliant data quality. It also directs our following survey assessment elements (See Section~\ref{sec:surveymethodology}).

The selected dimensions bridge several aspects of data quality: intrinsic properties (accuracy, completeness, bias mitigation), contextual and usability factors (relevance, accessibility, timeliness), and governance characteristics (documentation, traceability, security). Some dimensions were intentionally excluded to maintain a manageable scope. We retained only those dimensions exhibiting clear and significant overlap with regulatory requirements. Excluded dimensions either significantly overlap with broader categories (e.g., ``uniqueness'', ``reliability'', and ``consistency'', which correlate closely with accuracy) or receive comparatively less emphasis in the analysed legal frameworks.

\smallsection{Accuracy.} 
Accuracy means data correctness and reliability.
The GDPR explicitly establishes accuracy as fundamental, requiring personal data to be ``accurate and, where necessary, kept up to date'' (Art. 5(1)(d), Recital 39), and mandates rectification or erasure of inaccuracies (Art. 16). The AI Act similarly demands datasets for high-risk AI be, to the best extent possible, ``free of errors'' (Art. 10(3), Recital 67). Although both perspectives value accuracy highly, engineering typically pursues maximal precision quantitatively, whereas regulations adopt context-sensitive sufficiency standards rather than absolute accuracy.

\smallsection{Traceability.} 
Traceability involves tracking data changes, origins, and pathways within systems, enabling accountability and error analysis. The GDPR promotes traceability through mandatory records of data flows and modifications (Art. 30) and security logs (Art. 32). The AI Act requires automated event logging for high-risk AI systems (Art. 12) to support output-to-input accountability. Both technical and legal frameworks thus view traceability as essential for demonstrating compliance, clear attribution, and organisational responsibility.  

\smallsection{Relevancy.} Relevancy ensures data is appropriate and meaningful for its intended use. The GDPR’s data minimisation principle explicitly requires collected data to be ``relevant'' (Art. 5(1)(c)), while the AI Act similarly mandates dataset relevancy (Art. 10(3)), requiring explicit data suitability assessments (Art. 10(2)(e)). Data engineers often assess relevancy quantitatively, based on predictive value. The GDPR, however, applies a stricter qualitative standard of necessity and proportionality and restricts collection if alternatives to special categories of personal data exist. For compliance, teams must carefully plan data collection upfront, clearly justifying each variable legally and ethically beyond technical relevance alone.

\smallsection{Timeliness.} Timeliness means data is recently updated and available when needed, which is essential for accurate decision-making. The GDPR addresses timeliness within its accuracy principle, requiring personal data to remain ``up to date'' (Art.~5(1)(d)). While data engineers measure timeliness quantitatively (latency, data age) and may accept some degree of staleness in less time-sensitive data, legal standards are context-driven and risk-based, occasionally causing interpretive mismatches. Thus, organisations may need to define stricter update protocols than minimal legal requirements to ensure both regulatory compliance and data relevance.

\smallsection{Completeness.} Technically, completeness ensures that all necessary data elements are present, minimising missing values that could compromise analyses. Under the GDPR, completeness is required through data adequacy (Art. 5(1)(c)), requiring datasets to be sufficiently comprehensive for their purposes. The AI Act explicitly mandates dataset completeness as a development-stage requirement for high-risk AI systems (Art. 10(3)). Both technical and legal perspectives recognise completeness as crucial for reliability. However, there is tension with the GDPR's data minimisation principle, which requires that personal data processing is limited to what is strictly necessary. The GDPR also adopts a context-dependent, qualitative approach focusing on adequacy. Furthermore, the AI Act broadens completeness by including representativeness and coverage of populations, which broadens the traditional data engineering perspective that often focuses on per-record completeness.

\smallsection{Documentation.} Documentation entails clearly and comprehensively recording data, metadata, and management practices. The GDPR mandates thorough documentation via accountability (Art. 5(2), 35, Recital 82) and detailed records of processing activities (Art. 30(1), (3)\&(4)). Similarly, the AI Act requires extensive technical documentation for high-risk AI, detailing datasets and data management practices (Art. 11, Annex IV; Art. 13(3)(b)(vi), Art. 18, Annex XI). 
Technically, documentation primarily facilitates internal clarity and reproducibility. Legally, it serves as formal compliance evidence accessible to external stakeholders such as data subjects, auditors, or regulators. Documentation helps in translating complex technical activities into verifiable records to support both data quality management and regulatory obligations.

\smallsection{Accessibility.} 
Accessibility involves ensuring authorised users can easily retrieve and effectively use data. The GDPR reinforces accessibility via subject rights such as access (Art. 15) and portability (Art. 20). Thus, the GDPR's concept of accessibility centres around a user-centric model by encouraging the technical ability to facilitate transparency and control over users’ data. Both technical and legal perspectives recognise the balance between accessibility and security. This is typically achieved through role-based access controls and secure data-sharing methods in practice.

\smallsection{Security.} High-quality data management also includes confidentiality and security of personal information. The GDPR mandates ``integrity and confidentiality'' (Art. 5(1)(f)), requiring technical and organisational safeguards, privacy by design and by default (Art. 25), and risk-based security measures (Art. 32). Similarly, the AI Act emphasises privacy protection through the AI lifecycle (Recital 69), recommending secure methods like anonymisation and federated learning (Recital 27). Technically, data security methods like encryption and access control align directly with these regulatory standards. Regulations broaden data quality to include confidentiality and ethical usage. Notably, the security dimension is perceived differently between data quality literature and legal frameworks. From a legal standpoint, security mainly refers to technical and organisational safeguards preventing unauthorised access and ensuring data integrity, confidentiality, and availability (the CIA triad) \cite{Dhillon2001CurrentDI}. The technical literature often expands this dimension to include privacy-enhancing technologies and broader privacy protection measures.

\smallsection{Bias mitigation.} Bias mitigation refers to practices that identify, assess, and reduce unjustified systematic advantages or disadvantages across different groups. The GDPR explicitly prohibits automated decisions based on sensitive personal data (Art. 22(4), Recital 71). The AI Act requires datasets used in high-risk AI systems to proactively detect, prevent, and mitigate biases (Art. 10(2)(f,g); Art. 10(3); Art. 10(5), Recital 70, Recital 67). Although engineering typically balances fairness against predictive accuracy, legal frameworks impose additional societal fairness standards. This synergy encourages rigorous bias analysis and ethical data management.

\smallsection{Conclusion.} By systematically interpreting and operationalising technical definitions alongside relevant regulatory requirements, we have demonstrated significant alignment between data engineering best practices and legal mandates. Through highlighting potential points of divergence, this review also illustrates how practitioners might resolve conflicts between technical goals and regulatory constraints. Each core dimension of data quality represents a meeting point between technical practices and legal mandates.  Often, legal requirements encourage organisations towards existing engineering best practices. When differences emerge, they typically reflect the law’s attempt to balance competing societal values rather than fundamental disagreements on the importance of the dimension. 

Ultimately, the mapping presented here promotes interdisciplinary collaboration by offering a shared vocabulary and common understanding between data engineers, legal professionals, and policymakers. Such collaboration is essential to developing data practices that are both technically robust and compliant with regulatory standards, contributing to more responsible and effective data-driven systems.
However, our approach has a limitation: It focused exclusively on dimensions where clear alignment exists between technical practices and legal frameworks, omitting issues such as the legality of using third-party data for ML training and additional safeguards under the GDPR (e.g., Data Protection Impact Assessments, protection of special categories of personal data). Future research could expand on these considerations to explore broader legal contexts that impact data quality management.

%% file: 05-Survey.tex
\section{Survey Methodology}
\label{sec:surveymethodology}

\begin{figure}
\includegraphics[width=\columnwidth]{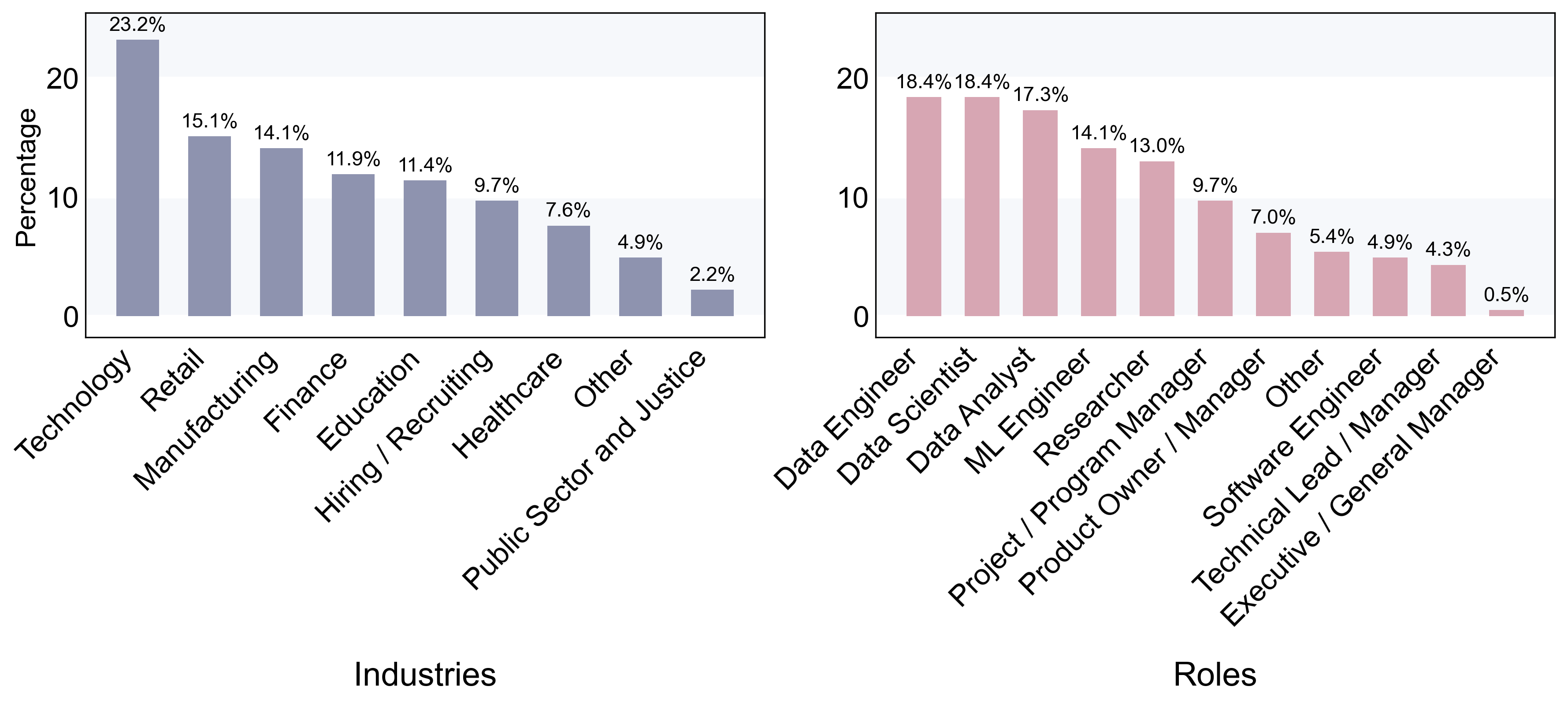}
  \Description{}
  \caption{Survey demographics: self-reported industries
(left) and team roles (right)}
  \label{fig:demographics}
\end{figure}

To better understand data practitioners' challenges and needs related to ensuring data quality in ML systems while aligning with regulatory requirements, we conducted an anonymous survey with 185 data practitioners from the EU. Here, we broadly defined "data practitioners" as individuals involved in any role within teams developing ML-based products or services. These roles range from operational roles (e.g., data collectors and labellers) and technical roles (e.g., data engineers, data scientists, and ML engineers) to design and management roles (e.g., product/project managers and general managers). The distribution of participants across these roles is illustrated in Fig.~\ref{fig:demographics}.

\subsection{Survey protocol}

Before designing the survey, we performed an extensive literature review (Section~\ref{sec:framework}). This review established a conceptual framework guiding the survey's assessment elements~\cite{Creswell2010ResearchDQ}. Specifically, the identified data quality dimensions were operationalised into structured answer options for questions concerning data quality management activities and challenges. These questions were explicitly framed around data quality work across the ML lifecycle, including data collection, preprocessing, training, deployment, and monitoring. By applying the framework to guide the formulation of survey items, we ensured a systematic mapping between theoretical constructs and empirical observations. 
The survey comprises three main sections: (1) demographics and background, (2) experience-based questions on finding and fixing data quality issues, and (3) needs and future expectations. A detailed list of questions is available in the supplemental material. For most questions, respondents could select multiple answers, presented in randomised order. 

A significant challenge in designing this survey was managing the sensitivity inherent in compliance-related topics. In compliance research, respondents might hesitate to answer truthfully due to concerns about social desirability, self-image preservation, or perceived risks \cite{tourangeau2007sensitive}. To mitigate this risk and encourage honest responses, we adopted several strategies. First, we clearly communicated that the survey would not collect personally identifiable or sensitive information, nor inquire about specific details related to participants' companies, products, or services. Secondly, we deliberately avoided normatively loaded questions, such as explicitly asking whether participants complied with or violated the regulations. In the experience-based section, the questions focused on factual descriptions of daily practices rather than direct references to rules or ethical judgements. Although response options implicitly aligned with the regulatory compliance requirements identified in Section~\ref{sec:related_work}, compliance obligations were never explicitly mentioned in the survey. Sensitive topics were intentionally integrated alongside non-sensitive topics. We employed neutral, technical language to minimise discomfort. By using neutral wording and avoiding judgmental or suggestive phrasing that could imply wrongdoing, the survey ensured respondents felt comfortable sharing truthful and reliable insights. This approach enabled the collection of rich, authentic data crucial for understanding the real-world challenges and needs of data practitioners. 

The survey was designed by a group of researchers specialising either in ML/AI or law. To validate and refine the survey design, we followed a three-stage iterative testing process. 
First, we conducted detailed case studies with two experienced data practitioners, each with over eight years of industry experience. During these case studies, we verbally asked and answered each survey question together, openly discussing potential misunderstandings and areas for improvement. Second, we carried out a pilot survey involving 15 internal researchers from our lab who had experience with ML projects. The primary goal of this pilot was to evaluate the functionality, logic, and user experience of the online survey interface. Finally, we convened a focus group composed of researchers and former data engineers. This group engaged in an in-depth discussion on the clarity, readability, and overall comprehensibility of survey instructions and questions. After each stage, we revised and improved the survey based on the feedback received. This iterative process significantly enhanced the quality and reliability of our survey instrument, ensuring close alignment with the core research objectives. The survey approach was reviewed and approved by the ethics committee of the authors’ institution.

\subsection{Survey implementation}
The finalised survey was conducted online using Qualtrics \cite{Qualtrics2025}. We recruited respondents using snowball sampling~\cite{Easterbrook2008SelectingEM}. Specifically, we distributed the survey link through emails and direct messages to our professional contacts across over 50 organisations in the EU. Recipients were encouraged to forward the survey to their colleagues involved in ML development, regardless of their specific roles. Additionally, we posted the survey through professional social networks and ML/AI-focused online communities, including LinkedIn, Discord, and Slack. 

A total of 250 participants started the survey. However, not all completed it. Reasons for incomplete responses likely include survey length (average completion time around 12 minutes during testing) and the technical complexity of the topic. During data analysis, we excluded responses completed too quickly or flagged as automated bots. To ensure data completeness and integrity, our analysis includes only the 185 respondents who finished all survey questions thoroughly. 

To enhance the representativeness regarding applicability to the GDPR and AI Act, we intentionally included questions targeting respondents whose teams process personal data (as defined by the GDPR) or develop products categorised as high-risk (under the AI Act, shown in Fig.~\ref{fig:demographics}). Additionally, respondents spanned 24 EU countries. The reported gender ratio (male to female, with additional gender options provided) was approximately 1.1, which reflects the known gender distribution within the ML industry \cite{young2023mind}. Nevertheless, we proactively attempted to mitigate potential gender imbalance by explicitly recruiting participants from female-focused communities, such as “Women in ML” groups. To ensure a broad representation, participants also varied considerably in their professional experience, industries, organisational sizes, and roles. Additional demographic details are provided in the supplemental material. While snowball sampling was effective in reaching a diverse set of respondents, it also introduces the potential for network-driven biases. The sample shows an overrepresentation of respondents from very large organisations with more than 1,000 employees (29.5\%). This skew may bias the results towards conditions typical of large enterprises, potentially underrepresenting perspectives from smaller organisations with more limited resources. The sample size similarly constrains subgroup analysis. We therefore examine patterns at the level of industry, organisational scale, and role, as finer cross-classifications would yield cells too small for reliable inference. Recruitment took place actively in November 2024 and continued until achieving at least 180 valid responses. At the start of the survey, all participants provided written informed consent and agreed to participate voluntarily without compensation. 

%% file: 06-Results_Discussion.tex
\section{Survey results}
\label{sec:results}

Our findings reveal that ensuring high data quality in ML systems that align with regulatory requirements (such as the GDPR and AI Act) presents several notable challenges and opportunities. Across the industry, practitioners identify missing data, documentation needs, and privacy considerations as key areas requiring attention in regulatory-aligned ML development. Notably, even organisations that actively apply mitigation measures for fairness and privacy continue to struggle with biased data and privacy risks, suggesting that current techniques are often insufficient to fully resolve these issues. These challenges point to unmet needs for more effective tools, refined practices, and better collaboration between ML practitioners and legal compliance teams to achieve robust data quality. In particular, gaps between the capabilities of existing data quality management tools and the features practitioners desire (such as automated validation and integrated compliance checks) underscore a significant need for improved tooling in this space. We organise the results by research questions, covering prioritised data quality aspects, lifecycle practices, tools in use and desired, and the role of legal–technical collaboration. To ensure that our findings are statistically reliable, we report proportions together with 95\% Wilson confidence intervals (95\% CIs) to show the likely range of population values. Correlations are accompanied by Fisher z-transformed CIs to indicate the stability of associations, while group differences are evaluated with the Mann–Whitney U test and summarised using Hodges–Lehmann median differences with 95\% CIs, which provide an interpretable effect size. 

\subsection{Prioritisation and best practices for data quality in regulatory contexts}


In addressing SRQ2, we investigated how practitioners prioritise and implement data quality dimensions in practice, especially when regulatory compliance strongly shapes their approaches. Data quality is widely understood as a multidimensional construct, with each category captured by specific quality dimensions. The activity options in the survey were derived from the data quality dimensions in our theoretical framework (Section~\protect\ref{sec:framework}), which capture both technical and legal relevance. Understanding which dimensions practitioners emphasise reveals how closely current practices align with compliance needs.

\begin{figure}
  \includegraphics[width=\columnwidth]{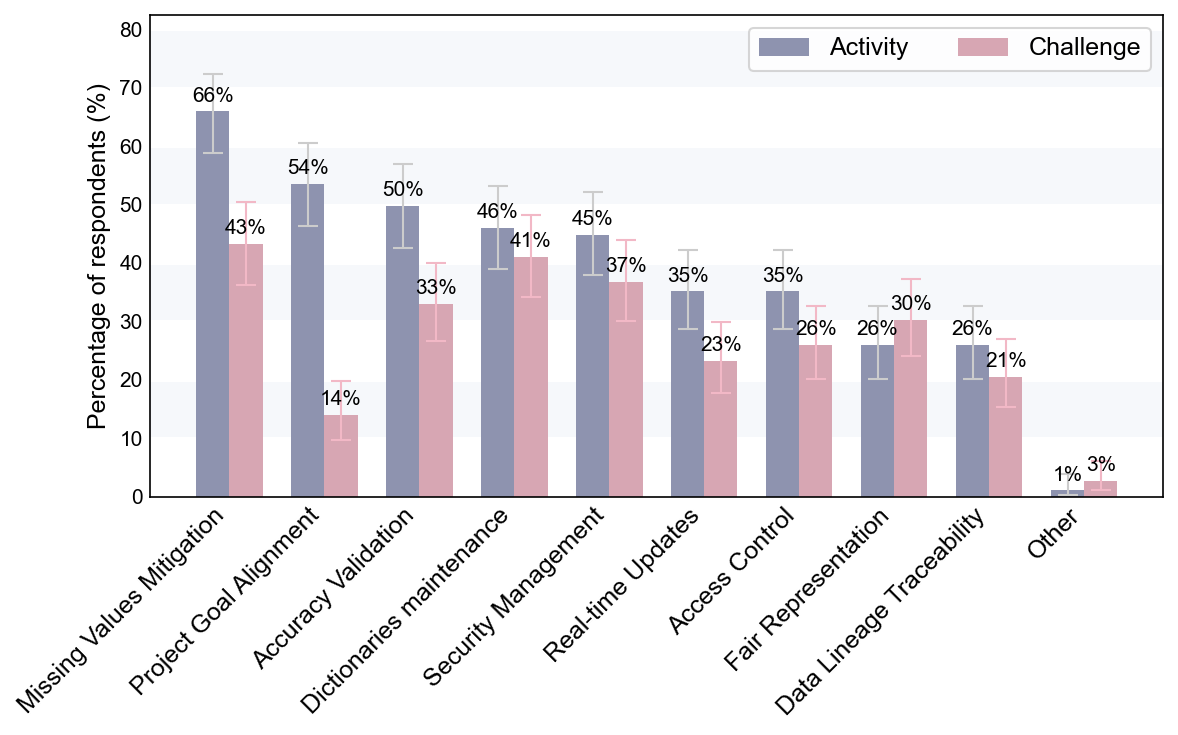}
  \Description{}
  \caption{Data quality management activities and challenges. Error bars indicate 95\% CI for the proportion of respondents.}
  \label{fig:activities}
\end{figure}

\smallsection{Core data quality priorities among practitioners (Fig.~\ref{fig:activities}).} Frequency analysis of multiple-choice responses shows that three data quality dimensions dominate practitioners’ attention, each chosen by more than half of respondents. The most prominent practice, "\textit{filling missing values}," was selected by over 66\% of respondents (95\% CI: 59–72\%), highlighting the centrality of completeness as a recurring obstacle in ML projects. Missing data directly undermines the reliability of analysis and model performance, and practitioners often resort to imputation or supplementary data collection to maintain analytical integrity. The prominence of this practice reflects both the ubiquity of incomplete datasets and the absence of fully satisfactory automated solutions.

The second most common practice, "\textit{aligning data attributes with project goals}," garnered 54\% of responses (95\% CI: 47–61\%). Practitioners place significant importance on ensuring that the data’s characteristics directly support the intended use case or business objective. This priority is driven by both operational and human-centred factors: aligning data with project goals improves model relevance and user satisfaction with the end system, and it facilitates better communication with stakeholders (e.g., project managers, domain experts) by using data that stakeholders consider meaningful. 

"\textit{Data verification for accuracy and reliability}" ranked third, selected by 50\% of respondents (95\% CI: 43–57\%). This includes checks like numerical range validation and format consistency to ensure data is correct and reliable. Interestingly, accuracy has long been considered a cornerstone of data quality, yet the survey indicates that teams often face practical limitations to achieve it. Our respondents note that exhaustive error checking or cleaning is resource-intensive, and in many cases accuracy controls are partial or selective. From a regulatory standpoint, inaccuracies in personal data could pose compliance issues with the GDPR, but practitioners seem to weigh the cost–benefit and often address what is feasible and reactive.

\smallsection{Multi-dimensional approach to data quality management.}
Analysis of selection patterns shows that most respondents selected between three and five practices, indicating a balanced yet focused approach to incorporating a range of data management dimensions within their projects. This predominant middle-ground strategy suggests a solid understanding of data quality's interconnected nature, where addressing one dimension often requires consideration of related aspects concurrently. Such an approach also reflects practical resource constraints that necessitate prioritising the most impactful combination of practices.

These selection patterns align with contemporary understanding of effective data management by acknowledging the need for balanced resource allocation across multiple quality dimensions. It reflects the growing alignment of what is deemed best data management practice in response to increasing data complexity, while demonstrating how organisations adapt their approaches based on specific contextual needs. The pattern suggests an industry-wide understanding that data quality cannot be ensured through single-point solutions, but rather requires a carefully orchestrated combination of complementary practices. They also contribute to our understanding that successful approaches typically involve a carefully balanced selection of complementary practices rather than an all-or-nothing approach.

\begin{figure}[t]
  \centering
  \includegraphics[width=\columnwidth]{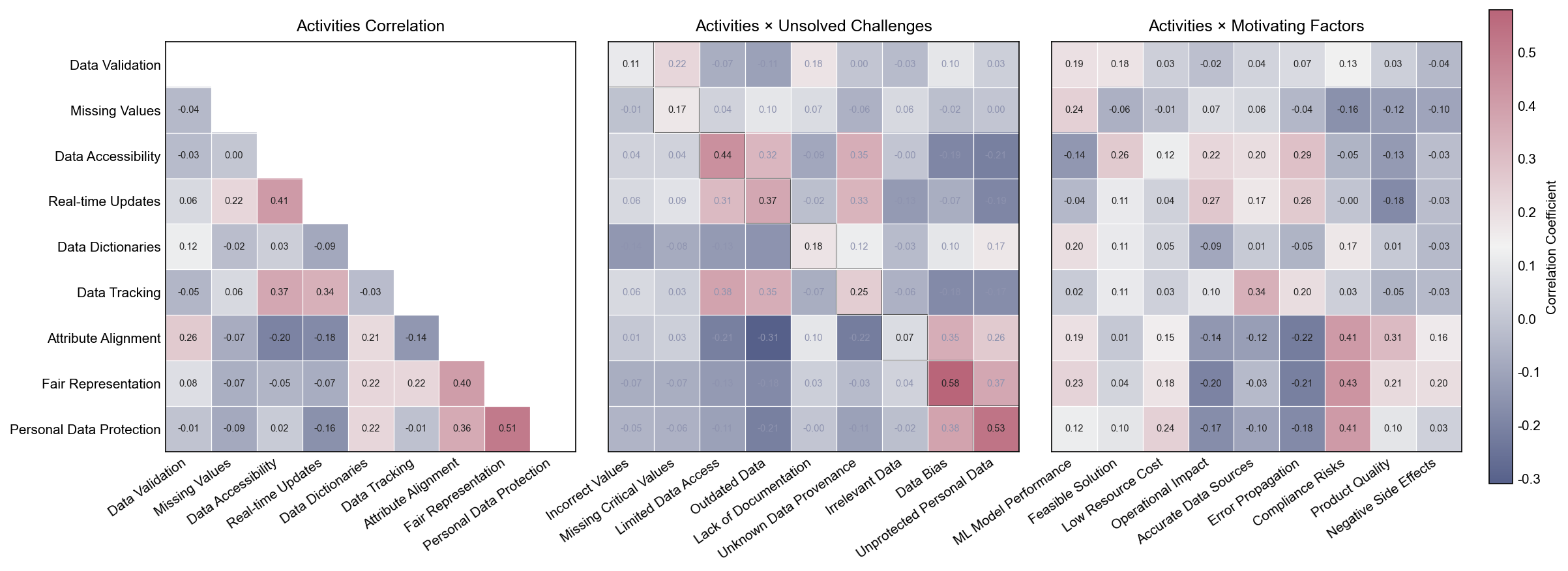}
  \Description{Three correlation heatmaps: left = activities;
  middle = activities vs. challenges;
  right = motivating factors vs. activities.}
  \caption{Correlation heatmaps: (a) Activities correlation; (b) Activities × unsolved challenges; (c) Activities × motivating factors.}
  \label{fig:activities_cor}
\end{figure}

\smallsection{Strategic alignments in data quality practices.}
Our analysis revealed significant patterns in how practitioners combine different data quality management practices (Fig.~\ref{fig:activities_cor}(a)). The correlations between various practices suggest both strategic alignments and emerging trends in how organisations approach data quality holistically.

The strongest correlation ($r \approx 0.51$, 95\% CI: 0.39–0.61, $p < 0.001$) emerged between \textit{ensuring fair representation} and \textit{implementing privacy-enhancing technologies (PETs)}. This relationship suggests that organisations recognising the importance of fairness in their data practices also tend to prioritise privacy protection. This correlation is particularly noteworthy in the context of regulatory frameworks that increasingly emphasise both algorithmic fairness and data privacy.

A moderate correlation ($r \approx 0.41$, 95\% CI: 0.28–0.52, $p < 0.001$) between \textit{implementing real-time data update mechanisms} and \textit{ensuring data accessibility} points to an emerging focus on dynamic data management systems. Organisations implementing real-time updates appear to recognise that the value of current data is maximised when it can be readily accessed by relevant stakeholders. This correlation suggests a strategic approach to data infrastructure that emphasises both temporal relevance and practical utility.

\smallsection{Key challenges in data quality management (Fig.~\ref{fig:activities}).}
Our survey revealed the challenges that practitioners face in ensuring data quality. The predominant challenge of \textit{missing critical values}, reported by 43\% of respondents (n = 80, 95\% CI: 36–50\%), reflects a fundamental obstacle in ML projects. This issue directly impacts model performance and reliability, potentially affecting compliance with regulations that require accurate and complete data. The second most reported issue was \textit{lack of documentation} (41\%, n = 76, 95\% CI: 34–48\%), followed by \textit{data privacy issues} (37\%, n = 68, 95\% CI: 30–44\%), \textit{incorrect values} (33\%, n = 61, 95\% CI: 27–40\%), and \textit{data bias} (30\%, n = 56, 95\% CI: 24–37\%). These challenges highlight the multifaceted nature of data quality concerns in regulatory contexts, where addressing one dimension often requires consideration of related aspects concurrently.

\smallsection{Persistent bias and privacy challenges despite active mitigation efforts (Fig.~\ref{fig:activities_cor}(b)).}
Our empirical investigation analysed the correlations between implemented data quality practices and persistent challenges to identify areas where current practices may be insufficient despite active management efforts.

Analysis revealed significant positive correlations between specific challenges and their corresponding mitigation activities. The \textit{Data Bias} challenge showed a strong positive correlation with \textit{Fair Representation Protocols} ($r \approx 0.58$, 95\% CI: 0.48–0.67, $p < 0.001$), while \textit{Privacy Protection} challenges demonstrated a similarly strong positive correlation with \textit{Privacy-Enhancing Technologies (PETs) Implementation} ($r \approx 0.53$, 95\% CI: 0.42–0.63, $p < 0.001$). These positive correlations suggest that organisations actively implementing these practices continue to encounter substantial challenges in both domains. 

We interpret these positive associations cautiously. They should not be read as evidence that fairness or privacy practices are ineffective. Rather, they likely indicate that teams adopt such practices precisely in contexts where the underlying problems are more visible, more severe, or more difficult to resolve. More broadly, the activities–challenge heatmap does not show a simple one-to-one relationship across all dimensions. Apart from the strongest same-domain associations, many relationships are relatively weak or diffuse, suggesting that data quality practices are often problem-driven and context-dependent rather than implemented as a uniform lifecycle-wide package. Bias and privacy issues are also structurally harder to close than isolated technical defects such as format errors or missing values: they depend on upstream data collection choices, organisational responsibility boundaries, domain-specific risk tolerance, and legal interpretation. This helps explain why mitigation activities and persistent challenges can co-exist. In practice, the presence of PETs or fair-representation protocols may signal greater awareness and more active management, while also revealing that available tools and procedures remain insufficient for fully resolving these issues.

\smallsection{Motivational factors driving data quality management (Fig.~\ref{fig:activities_cor}(c)).}
When asked about motivations for addressing data quality issues, the predominant motivating factor, cited by a significant majority of respondents, was \textit{the potential impact on model performance metrics} (68\%, n = 125, 95\% CI: 61–74\%). This finding aligns with the fundamental objectives of ML implementations, where model accuracy and effectiveness serve as primary indicators of project success. The prioritisation of performance-related factors suggests a pragmatic approach among practitioners, who recognise that data quality issues can propagate through the model pipeline. 

Our analysis of correlations between data quality activities and motivational factors reveals patterns to illuminate the primary drivers behind specific data quality practices in practitioners' approaches to data management. The practice of \textit{aligning data attributes with project goals} demonstrated strong positive correlations with two key motivators: \textit{overall product/service quality improvement} ($r \approx 0.31$, 95\% CI: 0.17--0.43, $p < 0.001$)  and \textit{regulatory compliance risk mitigation} ($r \approx 0.41$, 95\% CI: 0.28--0.52, $p < 0.001$). \textit{Fair representation practices for subgroups} exhibited significant positive correlations with similar motivational factors: \textit{product/service quality enhancement} ($r \approx 0.21$, 95\% CI: 0.07--0.34, $p < 0.01$) and \textit{regulatory compliance} ($r \approx 0.43$, 95\% CI: 0.30--0.54, $p < 0.001$). These correlations indicate that demographic fairness initiatives are substantially driven by both quality improvement objectives and regulatory requirements.

\smallsection{Stage-specific data quality needs.}
Our analysis reflects how data quality practices vary across different stages of the ML lifecycle. A key feature of the analysis is its focus on specific data quality dimensions, which recur at multiple stages but with differing priorities.
A particularly significant observation is the evolution of quality priorities across stages. For example, during the \emph{dataset design and collection} phase, \textit{relevance} (n = 86) is prioritised, emphasising the importance of ensuring data suitability for intended use cases. Additionally, the focus on \textit{accuracy} during this stage supports the literature's view that a robust foundation of intrinsic data quality is essential for effective modelling and analysis later on. Mistakes or shortcomings at this stage can have cascading effects, compounding issues throughout the project.

The \emph{data processing} phase stands out with a substantial 380 recorded instances of quality-related activities, underlining previous assertions about the critical role of data validation and maintenance during this phase. This stage also sees a shift towards emphasising \textit{completeness} (n = 114) and maintains a strong focus on \textit{protecting personal information}. Crucially, the processing stage is when many regulatory compliance measures are implemented on the data. Our findings show a strong focus on protecting personal information during data processing. Best practices here include applying privacy-enhancing techniques such as anonymisation or pseudonymisation of personal identifiers, aggregating data to a less sensitive form when possible, and enforcing access controls on the raw data. 

Once \emph{models are trained and deployed}, data quality efforts shift towards maintaining quality over time and in production. Best practices in this stage revolve around ensuring \textit{timeliness} (n = 46), \textit{accessibility} (n = 18), and continued \textit{validity of data}. Ensuring this often involves setting up proper data architecture: using databases or data lakes with well-managed access permissions, indexing data for quick queries, and providing documentation on where the model’s input data resides. This practice aligns with regulatory needs for accountability, where an organisation might be asked to explain a model’s decision by showing the input data and how it was processed. It also connects to user-facing aspects. For example, if a user requests their data under the GDPR’s right of access, the company should be able to easily retrieve all relevant data, which is facilitated by good data accessibility practices.

The analysis also highlights consistent considerations of \textit{privacy} and \textit{fairness} across all lifecycle stages, with \textit{privacy} concerns noted in 69, 40, and 35 instances in data processing, ML monitoring, and ML evaluation phases, respectively, and \textit{fairness} in 23, 34, and 30 instances in data collection, data processing, and ML evaluation phases. This indicates a continuous effort to address potential biases and ensure equitable outcomes, resonating with the literature's emphasis on the importance of fairness in data handling and algorithm training to prevent biases that could lead to unfair model outcomes. Privacy, too, is a critical ongoing concern across multiple stages, especially given the increasing focus on data protection regulations.

Throughout all stages, a unifying best practice is cross-stage consistency and handoff. This means information gained at one stage is propagated to the next. For instance, if during data collection it was noted that certain sensitive attributes are present, that knowledge should inform processing (to apply privacy treatments) and deployment (to monitor how those attributes might affect outcomes). Drawing on empirical evidence from our survey, we argue that data quality management cannot be “one and done.” Therefore, a comprehensive, stage-aware approach is considered best practice for aligning ML data quality with regulatory standards.

\smallsection{Role-specific contributions to data quality.}
The distribution of roles across different phases of the ML lifecycle indicates a strategic positioning and varied contributions at each stage. \textit{Data analysts} are consistently present throughout the lifecycle, acting as crucial connectors between different stages and stakeholders. Their ongoing role underscores the importance of continuous monitoring and management of data quality, particularly crucial as data contexts and characteristics can change unpredictably. \textit{Data scientists} demonstrate a notable increase in involvement in the later stages of the pipeline, where their skills in model development and evaluation become indispensable. However, their less frequent involvement in the early stages suggests a potential underutilisation of their skills in initial data quality assessments, where their input could be highly beneficial. \textit{Data engineers} (n = 13) are primarily active in the data processing and infrastructure development phases. Their focus reflects their essential role in constructing and maintaining the technical frameworks that support high data quality, including the development of robust data pipelines and the automation of quality checks.

Highlighting the importance of cross-functional collaboration, the analysis suggests a high degree of cooperation among roles in managing data quality. The lifecycle view shows that managing data quality in a regulated environment isn’t a single-step fix, but a continuous, collaborative process that involves many stakeholders and tools at different points in time.

\subsection{Data quality management tools: current usage and future needs}

\smallsection{Current tool utilisation.}
\begin{figure}
\includegraphics[width=\columnwidth]{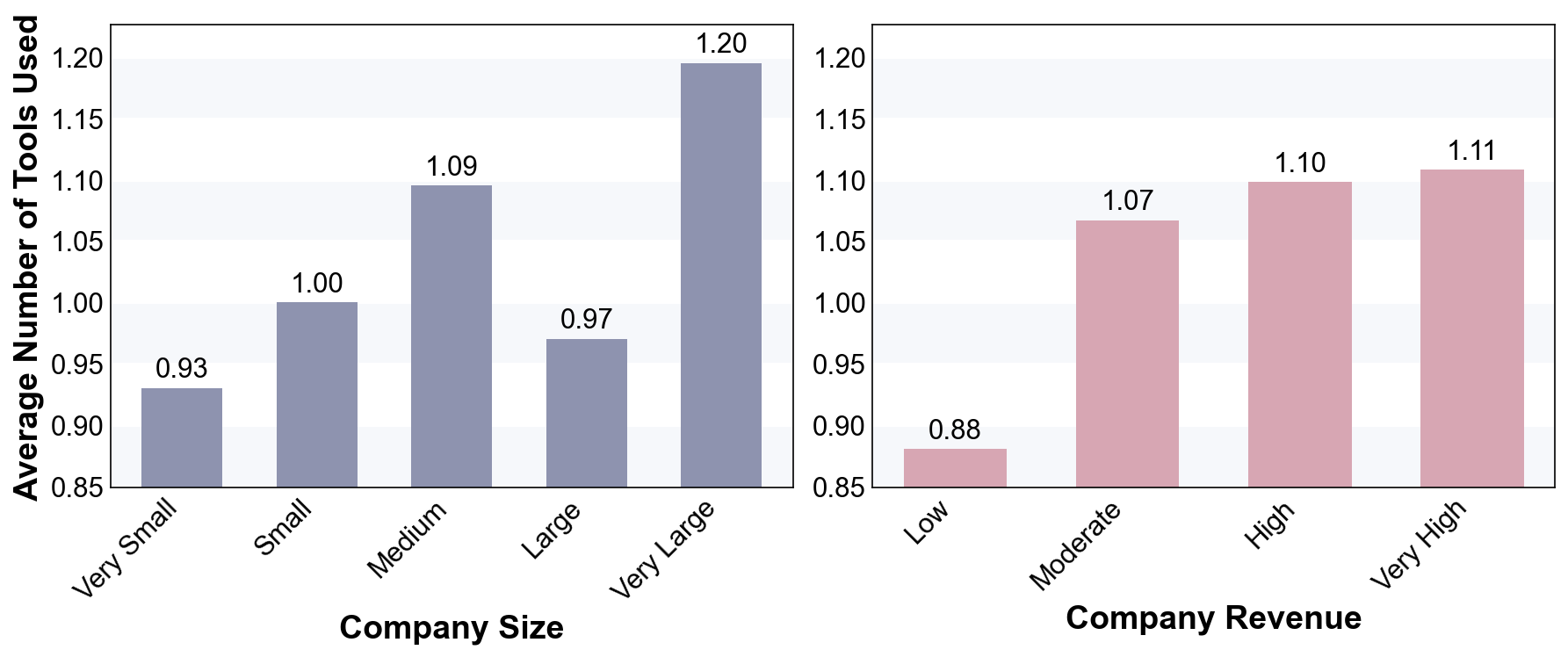}
  \Description{}
  \caption{Organisational scale and tool diversity. Bars show the average number of tool categories selected by respondents in each organisational size and revenue group.}
  \label{fig:orgtool}
\end{figure}
Addressing SRQ3, we investigated the tools and methodologies practitioners currently employ to manage data quality. Our empirical analysis reveals a significant reliance on custom solutions among data practitioners, with 22\% employing \textit{bespoke scripts and in-house tools}. Respondents highlighted several advantages of using in-house tools, noting that "In-house tools give us full control and are tailored for specific use cases". For example, one respondent mentioned using a "self-developed AI-based data cleaning tool that performs sophisticated tasks without heavy feature engineering". Another advantage cited was that "(custom tools can be) optimised for internal applications", suggesting a potential gap in the market where standardised solutions fail to meet specific organisational needs.

Among commercial and open-source options, \textit{Azure Data Factory} and \textit{TensorFlow Data Validation (TFDV)} are the most adopted, particularly by enterprise-scale organisations that leverage cloud infrastructure. Fig.~\ref{fig:orgtool} suggests that respondents from larger and higher-revenue organisations reported using slightly more tool categories on average. However, the absolute differences are small, so we treat this result as suggestive rather than as strong evidence of a substantial organisational-scale effect.

\smallsection{Future tool needs.}
Survey responses about desired tool features (averaging 3.0 features per respondent) revealed a substantial divergence between current practices and desired capabilities. It highlights significant opportunities for tool evolution and market development. The most sought-after feature was \textit{automated data validation} (n = 69), reflecting a strong market demand for sophisticated 
validation frameworks. This preference likely stems from the resource-intensive nature of current manual validation processes.

\textit{Regulatory compliance frameworks} also drew significant interest (n = 68), underscoring the high demands of adhering to evolving regulations. Similarly, \textit{privacy protection mechanisms} were highly valued (n = 58), along with substantial demand for \textit{bias detection systems} (n = 51) and \textit{security implementation features} (n = 45). The concentration of interest in governance-related features indicates an increasing awareness of the regulatory complexities in data management practices.

A notable insight came from respondents who currently rely on custom-built solutions for data quality. Even among these practitioners who have tailored systems, 12.3\% of them explicitly voiced a need for \textit{comprehensive documentation capabilities} in tools. This indicates an urgent need to formalise institutional knowledge and standardise processes, especially in environments relying on bespoke solutions.

Indeed, our analysis identified a substantial gap between what current tools offer and what practitioners say they want. Taken together, these needs define a vision for future data quality tools: integrated platforms that combine functions such as automated data validation, compliance auditing, privacy and bias mitigation, and thorough documentation. Most existing solutions handle only pieces of the puzzle. Practitioners currently juggle these disjointed tools or fill gaps with manual effort. An ideal solution would seamlessly integrate multiple functions and substantially reduce human burden and help standardise best practices. Better user-centred design of data quality management tools is needed. These designs should focus on workflow integration so that quality checks and compliance steps fit naturally into the ML development process. Usability should also be prioritised, as many of these practitioners may lack expertise in legal compliance or advanced statistics. Tools should surface potential data issues intuitively and guide users on how to fix them. This approach effectively bridges the knowledge gap between what data engineers know and what legal requirements demand.

\begin{figure}
\includegraphics[width=\columnwidth]{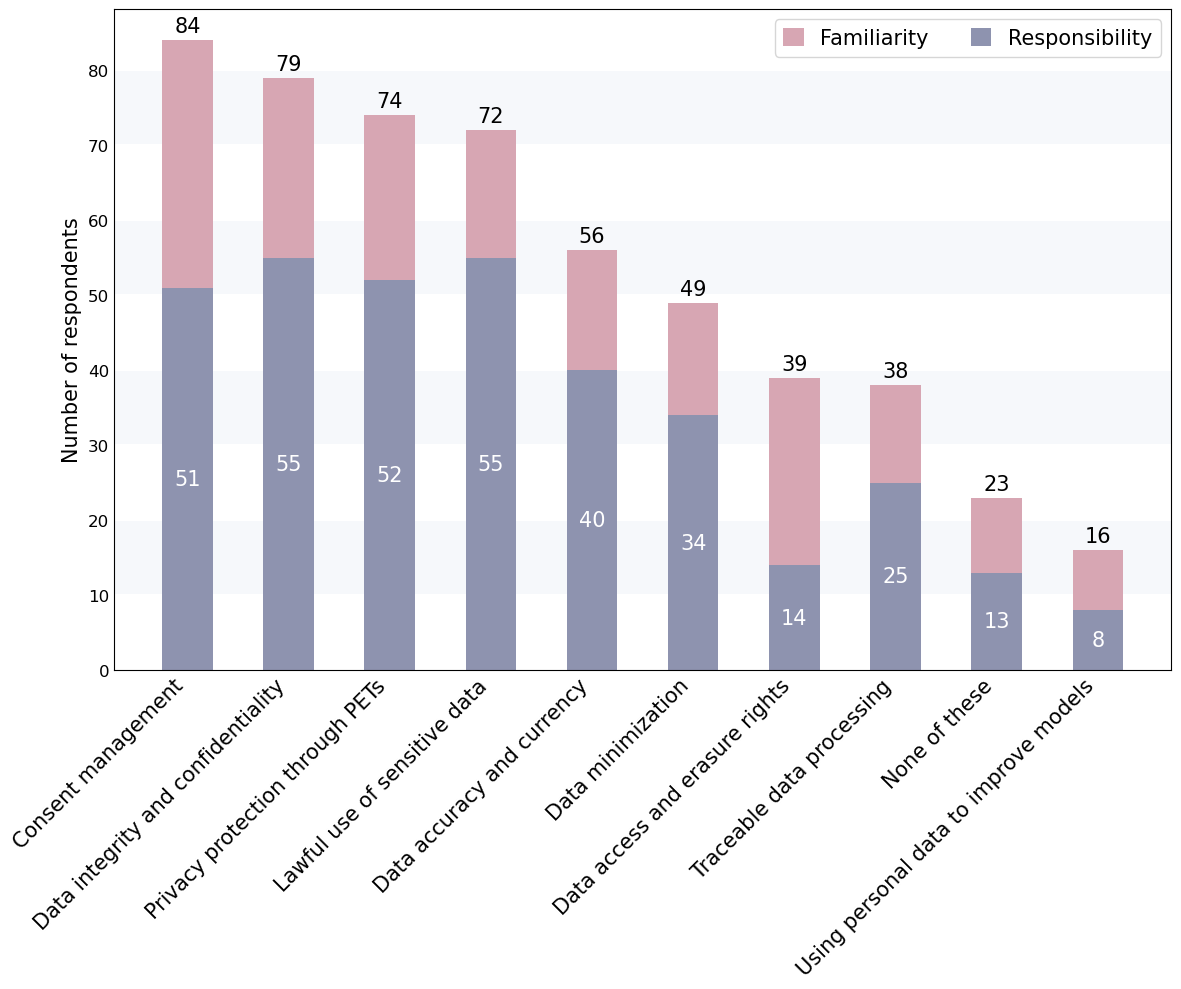}
  \Description{}
  \caption{Familiarity and responsibility}
  \label{fig:familiarity}
\end{figure}

\smallsection{Cross-functional collaboration between data and legal teams.}
In our survey, we asked practitioners about their \textit{familiarity} with regulatory aspects of personal data processing and their perceived \textit{responsibility} for implementing them (Fig. \ref{fig:familiarity}). One of the clearest findings is the difference between practitioners who collaborate with legal teams and those who do not. Data practitioners who reported \textit{collaborating with legal teams} demonstrated markedly higher engagement with regulatory aspects than those without such collaboration. In terms of familiarity, collaborators knew on average more regulatory aspects (mean number of aspects, M = 3.81) than non-collaborators (M = 1.94). This difference was statistically significant (Mann--Whitney U = 1971.5, $p < 0.001$) and corresponded to a Hodges--Lehmann median difference of 2.0 aspects (95\% CI: 1–3). A similar pattern emerged for responsibility: collaborators reported responsibility for more aspects on average (M = 2.40) compared to non-collaborators (M = 1.26; Mann–Whitney U = 2013.5, $p < 0.001$), again with a median difference of roughly two aspects (95\% CI: 1–3). 

These results indicate that legal--technical collaboration is associated with broader regulatory awareness and stronger ownership of compliance-related tasks. Working with legal teams equips data practitioners to anticipate regulatory requirements earlier in the pipeline, reducing reliance on reactive reviews. It also provides a shared vocabulary that helps align technical practices with legal expectations, making joint assessments of risks such as fairness, provenance, or privacy more systematic. In this sense, cross-functional collaboration is not simply supportive but transformative, embedding compliance as a routine element of data quality management and fostering more trustworthy ML development.

When examining specific personal data protection aspects, we found that \textit{consent management} was the most familiar aspect (57\%), followed closely by \textit{integrity and confidentiality of personal data} (53\%). These were also the areas where practitioners felt most responsible (40\% and 43\% respectively). These familiarity figures are computed over respondents whose teams process personal data (n = 148), while responsibility is asked only of those familiar with an aspect (n = 127). While 50\% of these respondents were familiar with \textit{privacy-enhancing technologies}, only 41\% of those familiar felt responsible for implementing them, indicating a potential gap between knowledge and implementation ownership.

\textit{Industry} context also modulates this collaboration impact. Practitioners in heavily regulated sectors such as \textit{healthcare (4.5 aspects on average), the public sector and justice (4.4 aspects) and hiring/recruiting (4.1 aspects)} reported much higher levels of engagement with personal data management. In contrast, industries like \textit{technology (2.4), retail (2.1), and manufacturing (1.8)} were noted as having comparatively lower involvement in these practices. The implication is that where cross-functional dialogue is lacking, even aware practitioners might not implement certain quality measures due to uncertainty or deprioritisation.

Another dimension is the \textit{organisational scale}. We observed a clear positive correlation between the size (and financial robustness) of a company and the level of data compliance engagement by practitioners. In settings where legal guidance is readily available (as is often the case in big firms), compliance becomes part of the culture and workflow, whereas where it’s absent, practitioners may unintentionally neglect important practices. Our results support strategies like providing legal support mechanisms for smaller companies and start-ups, so that they can elevate their data practices.

Despite the clear benefits of cross-functional collaboration, our research identified persistent challenges including: (1) gaps between privacy technology familiarity and implementation responsibility, (2) continued documentation shortfalls despite improved awareness, and (3) predominantly reactive rather than proactive collaboration patterns. These gaps suggest opportunities for implementing emerging best practices such as integrated project teams, shared technical-legal vocabularies, and compliance-as-code approaches that embed regulatory requirements directly into validation frameworks.

These findings suggest organisations should create formal structures for ongoing dialogue between legal and technical teams throughout the ML lifecycle, invest in developing professionals with hybrid expertise, and implement earlier legal involvement in technical planning to avoid costly redesign when regulatory issues emerge late in development. This human collaboration aspect is a reminder that data quality is not just a technical issue, but a socio-technical one: the interplay of roles and communication channels directly shapes how well data quality challenges are addressed in practice. As regulations evolve, this interplay will become even more pivotal.

%% file: 07-Recommendations.tex
\section{Recommendations and Future Directions}
\label{sec:recommendations}
Building on our survey findings and theoretical framework, we first illustrate how the proposed framework can be applied in practice, and then propose actionable recommendations for practitioners under two themes alongside several directions for future research.
Together, these steps illustrate how organisations and researchers can advance towards integrated, compliance-aware data quality management.  

\smallsection{Applying the Framework in Practice} The framework can be used as a lightweight lifecycle checklist. Consider a CV-screening system, which qualifies as high-risk under the AI Act. At project scoping, the team documents the intended purpose, lawful basis, data subjects, and the data quality dimensions expected to matter. During collection and preprocessing, completeness, accuracy, relevance, and data minimisation are assessed together, so that missing application records or feature choices are not treated as purely technical concerns. During training and evaluation, the team verifies that subgroup performance, representativeness, and bias checks are documented (AI Act Art. 10). Before deployment, it confirms that data lineage, access controls, update procedures, and audit evidence are in place. After deployment, monitoring covers not only model drift, but also whether the data remains timely, relevant, and consistent with the documented purpose. Read in this way, Table 1 functions as a stage-by-stage guide rather than a static mapping.

\smallsection{Tool Integration and Automation} 
The empirical findings (Section~\ref{sec:results}) underscore the need for more integrated data quality tools. Our framework in Section \ref{sec:framework} provides a blueprint for such integration. We recommend compliance-aware data quality contracts that require each dataset or feature to declare the lawful basis, purpose, subject categories, retention, sensitive attributes, transfer restrictions, and measurable targets. Dashboards should track quality dimensions (accuracy, completeness, timeliness, etc.) alongside compliance indicators, with legal metadata propagated through attribute-level lineage. Documentation warrants particular attention. Where it is reconstructed after the fact, it is costly and often incomplete. Tools should instead capture evidence as a by-product of normal pipeline execution. Binding these contracts to pipeline gates ensures deployments fail if quality or compliance checks regress, preventing unlawful use and simplifying audits. Such gates work best when placed early. Late interventions such as drift detection or fairness-aware retraining are corrective by nature and cannot undo choices made at collection. Bias and privacy checks should therefore be shifted into dataset design, where representativeness and lawful basis can still be adjusted at reasonable cost.

\smallsection{Cross-Functional Collaboration and Hybrid Expertise}
The survey findings highlight the critical need for better collaboration between technical and legal practitioners. Where this collaboration is structural rather than ad hoc, practitioners engage with regulatory requirements far more actively. Assigning a named legal contact to each ML project is a low-cost first step, as it allows questions to be raised without formal escalation. Our framework can serve as a shared vocabulary for proactive planning rather than reactive review. It can be used to establish internal training programmes that educate data practitioners on the nuances of GDPR and AI Act, and legal teams on the technical challenges of data quality in ML. Mandatory joint review checkpoints at key lifecycle stages should enable legal, ethics, and technical teams to collectively assess data quality, compliance risks, and mitigation strategies. For example, prior to model deployment, a joint review could verify that fairness metrics meet defined thresholds and that data provenance is fully documented. Collaboration should be paired with clear ownership. Naming an owner for each quality dimension, whether an individual or a team, prevents recognised issues from remaining unclaimed. A practical step is to maintain a shared, version-controlled repository of compliance rules expressed in a domain-specific language or configuration format. This can reduce misinterpretations and foster more effective problem-solving.

\smallsection{Future Research Directions}  
To extend these recommendations, future research will build on our theoretical framework and survey findings through an interview study with industry practitioners. This qualitative phase will allow us to examine in greater depth how organisations operationalise compliance-aware data quality practices, and to uncover the social, organisational, and technical barriers that may hinder adoption. In particular, we will investigate how practitioners negotiate trade-offs between legal obligations and engineering constraints, how collaboration with compliance teams is initiated in practice, and how data quality tools are adapted for compliance tasks.

By systematically adopting these strategies and exploring these research avenues, organisations and researchers can move towards a state where data quality management inherently aligns with EU regulatory requirements, transforming compliance from a reactive overhead into an integral component of responsible and trustworthy ML development and deployment.

%% file: 08-Conclusion_Future.tex
\section{Conclusion}
\label{sec:conclusion}

In conclusion, our research highlights the complex interplay between technical data quality needs and regulatory compliance requirements in ML systems. There is a clear emergence of compliance-driven practices as essential components of data quality management. Rather than viewing regulatory compliance as an additional demand, practitioners could significantly benefit from methodologies that seamlessly integrate regulatory requirements into their technical practices. 

Although our legal mapping is grounded in the GDPR and the EU AI Act, the broader findings may also be relevant beyond the EU. The specific legal obligations will differ across jurisdictions, but practitioners in other contexts may face similar challenges in translating governance requirements into data collection, preprocessing, documentation, access control, monitoring, and audit practices. For example, in more sector-specific regulatory environments such as the United States, the exact mapping would need to be re-examined, while the organisational challenges of fragmented tooling, unclear ownership, and limited legal-technical collaboration may still apply. The GDPR in particular has served as a blueprint for data protection laws in many jurisdictions, so its core principles of accuracy, minimisation, and accountability have close counterparts elsewhere, for example in California under the CCPA and CPRA. The AI Act has fewer direct analogues, since the EU is an early mover in statutory AI regulation. A related interview study of EU practitioners~\cite{wang2026detective} finds the same implementation gaps recurring across organisations, which suggests these challenges arise from the general difficulty of operationalising governance requirements rather than from any single legal regime.

Solid data quality practice will also benefit regulatory compliance; however, it does not guarantee full compliance. By situating these insights within our theoretical framework, we demonstrate how legal and technical perspectives can be systematically connected, laying the groundwork for more robust and trustworthy ML development in regulated contexts.